\documentclass[english,twocolumn,showpacs,preprintnumbers,amsmath,amssymb]{revtex4}

\usepackage{graphicx}
\usepackage{babel}

\begin{document}

\title{Breaking rate minimum predicts the collapse point of over-loaded materials}

\author{Srutarshi Pradhan}
\email{srutarshi.pradhan@sintef.no}
\affiliation{SINTEF Petroleum Research, NO-7465 Trondheim, Norway}

\author{Per C. Hemmer}
\email{per.hemmer@ntnu.no}
\affiliation{Department of Physics, Norwegian University of Science and
Technology, NO-7491 Trondheim, Norway}

\begin{abstract}
As a model of composite materials, we choose a bundle of fibers with 
stochastically distributed breaking thresholds for the individual fibers.
The fibers are assumed to share the load equally, and to obey Hookean elasticity
right up to the breaking point. We study the evolution of the fiber breaking rate
at a constant load in excess of the critical load. The analysis shows that  
the breaking rate reaches a minimum when the system is half-way from 
its complete collapse. \\

\end{abstract}


\maketitle

\section{Introduction}
Bundles of fibers, with statistical distributed thresholds for the breakdown of 
individual fibers, present interesting models of
failures in materials. They have simple geometry and clear-cut rules for how 
stress caused by a failed element is redistributed on 
undamaged fibers.  Since these models can be analyzed to an extent that is not possible for more complex materials, they have been much studied 
(For reviews, see \cite{Herrmann,Chakrabarti,Sornette,Sahimi,Bhattacharyya}). 
The statistical distribution of the \textit{size} of avalanches in fiber
bundles is well studied \cite{HH,PHH05,HHP}, and the failure 
dynamics under constant load has been formulated \cite{PBC}  through recursion
 relations which in turn explore the phase transitions and associated critical 
behavior in these models. 
 
In this article we present a way to predict when an 
over-loaded bundle collapses, by monitoring the fiber breaking rate.

We focus on the equal-load-sharing models, in which the load previously carried
by a failed fiber is shared equally by all the remaining intact fibers
\cite{Peirce,Daniels,Smith,Phoenix}. We consider a
bundle consisting of a large number $N$ of elastic fibers, clamped
at both ends (Fig.\ 1). The fibers obey Hooke's law with force constant set 
to unity for simplicity.
Each fiber $i$ is associated with a breakdown threshold $x_{i}$
for its elongation. When the length exceeds $x_{i}$ the fiber breaks
immediately, and does not contribute to the strength of the bundle
thereafter. The individual thresholds $x_{i}$ are assumed to be independent
random variables with the same cumulative distribution function $P(x)$
and a corresponding density function $p(x)$: \begin{equation}
{\rm Prob}(x_{i}<x)=P(x)=\int_{0}^{x}p(y)\; dy.\end{equation}
\begin{center}
\includegraphics[width=6cm,height=6cm]{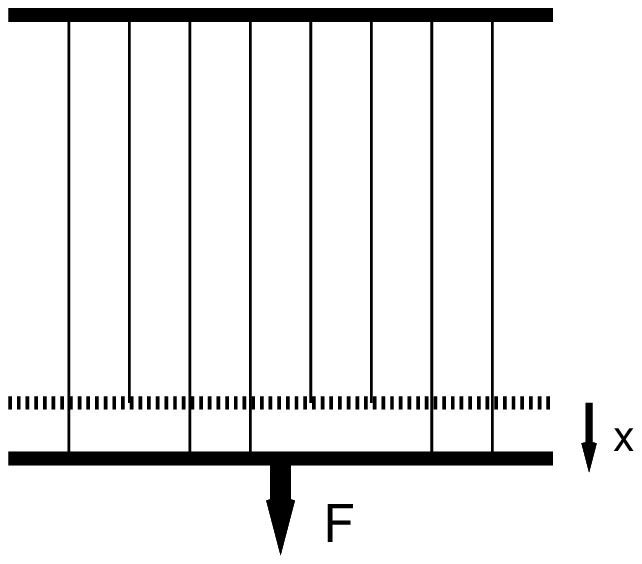}
\par\end{center}

\noindent {\small FIG.\ 1. The fiber bundle model. }\\
{\small \par}

If an external load $F$ is applied to a fiber bundle, the resulting
failure events can be seen as a sequential process \cite{PBC}.
In the first step all fibers that cannot withstand the applied load
break. Then the stress is redistributed on the surviving fibers, which
compels further fibers to fail, etc. This iterative process continues
until all fibers fail, or an equilibrium situation with a nonzero
bundle strength is reached. Since the number of fibers is finite,
the number of steps, $t_{f}$, in this sequential process is {\em
finite}. 

At a force (or elongation) $x$ per surviving fiber the total force on 
the bundle
is $x$ times the number of \textit{intact} fibers. The expected or average
force at this stage is therefore \begin{equation}
F(x)=N\, x\,(1-P(x)).\label{load}\end{equation}
The maximum $F_{c}$ of
$F(x)$ corresponds to the value $x_{c}$ for which $dF/dx$ vanishes.
Thus \begin{equation}
1-P(x_{c})-x_{c}p(x_{c})=0.\end{equation}
We characterize the state of the bundle as \textit{pre-critical} or
\textit{post-critical} depending upon the stress value $\sigma=F/N$ relative to
the critical stress \begin{equation}
\sigma_{c}=F_{c}/N,\end{equation}
We study the stepwise failure process in the bundle, when a fixed
external load $F=N\sigma$ is applied. Let $N_{t}$ be the number
of intact fibers at step no.\ $t$, with $N_{0}=N$. We want to determine
how $N_{t}$ decreases until the degradation process stops. With $N_{t}$
intact fibers, an expected number \begin{equation}
\left[NP(N\sigma/N_{t})\right]\end{equation}
 of fibers will have thresholds that cannot withstand the load, and
consequently these fibers break immediately. Here $[X]$ denotes the
largest integer not exceeding $X$. The number of intact fibers in
the next step is therefore
 \begin{equation}
N_{t+1}=N-\left[NP(N\sigma/N_{t})\right].\label{Nt}\end{equation}
 Since $N$ is a large number, the ratio \begin{equation}
n_{t}=\frac{N_{t}}{N}\end{equation}
 can for most purposes be considered a continuous variable. By (\ref{Nt})
we have essentially \cite{PBC} \begin{equation}
n_{t+1}=1-P(\sigma/n_{t}).\label{n}\end{equation}

\section{ The relation between minimum breaking rate and complete collapse}

We will now demonstrate, for three different threshold distributions, that 
there is a a relation between the minimum of the breaking rate 
$R(t)=-dn_t/dt$ (treating $t$ as continuous) and the moment $t_f$ when the 
complete fiber bundle collapses.   

\subsection{ Uniform distribution}
We consider the uniform distribution, $P(x)=x$ for $0\leq x\leq 1$, and assume 
that the load is post-critical:
$ \sigma = \frac{1}{4} +\epsilon,$
with $\epsilon>0$. Simulations show that the breaking rate has a  
minimum at some value $t_0(\epsilon)$, and that for varying $\epsilon$  the 
minima all occur at a value close to $\frac{1}{2}$ when plotted as function 
of the scaled variable $t/t_f$ (Fig.\ 2).\\ 

\begin{center}
\includegraphics[width=6cm,height=6cm]{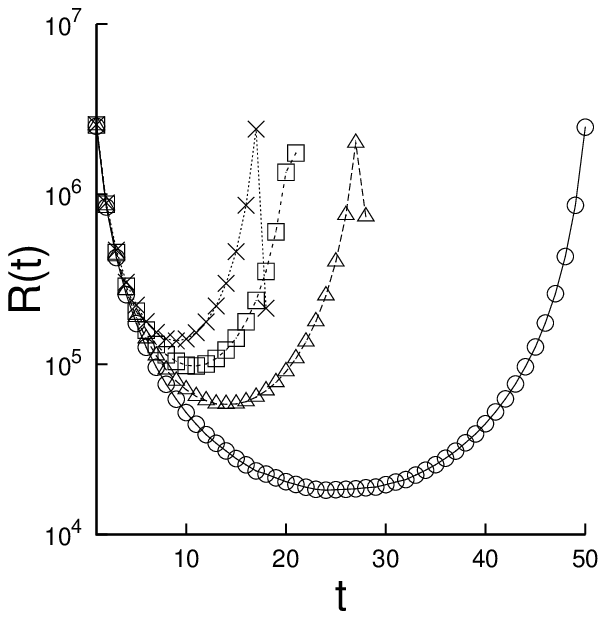}
\includegraphics[width=6cm,height=6cm]{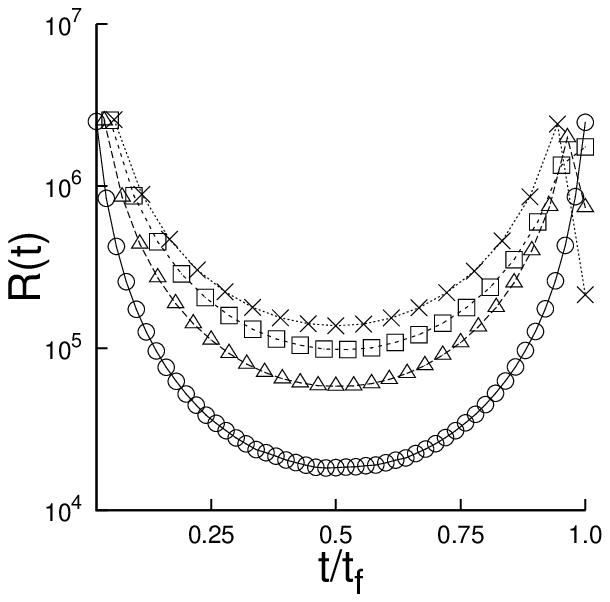}
\par\end{center}

\noindent {\small FIG.\ 2. The breaking rate $R(t)$ vs. step $t$ (upper plot) 
and vs.\ the rescaled step variable $t_f/t$ (lower plot) for the uniform 
threshold distribution for a bundle of $N= 10^7$ fibers. 
Different symbols are used for different excess stress levels 
$\sigma -\sigma _c$: 0.001  (circles), 0.003 (triangles), 0.005 (squares) 
and 0.007 (crosses). }\\{\small \par}

This can be shown analytically. The iteration (\ref{n}) takes in this case the 
form
\begin{equation}
n_{t+1} = {\textstyle \frac{1}{2}} - ({\textstyle \frac{1}{4}}+\epsilon)\frac{1}{n_t}.
\label{unin}\end{equation}
By direct insertion one verifies that
\begin{equation}
n_t =  {\textstyle \frac{1}{2}} -\sqrt{\epsilon}\tan (At-B),
\label{un}\end{equation}
where
\begin{equation}
 A= \tan^{-1}(2\sqrt{\epsilon})\hspace{1cm}\mbox{and}\hspace{1cm}B=\tan^{-1}(1/2\sqrt{\epsilon}),
\end{equation}
is the solution (\ref{unin})satisfying the initial condition $n_0=1$. From (\ref{un}) follows
the breaking rate
\begin{equation}
R(t)=- \frac{dn_t}{dt} = \sqrt{\epsilon}A\cos^{-2}(At-B).
\end{equation}

$R(t)$has a minimum when 
\begin{equation}
0=\frac{dR}{dt} \propto \sin(2At-2B),
\end{equation}
which corresponds to
\begin{equation}
t_0=\frac{B}{A}.
\end{equation}
When criticality is approached, i.e.\ when $\epsilon \rightarrow 0$, we have $A\rightarrow 0$, and thus $t_0\rightarrow \infty$, as expected.

We see from eq.(\ref{un}) that $n_t=0$ for 
\begin{equation}
t_f = \left(B+\tan^{-1}(1/2\sqrt{\epsilon}\right)/A= 2B/A.
\end{equation}
This is an excellent approximation to the integer value at which the fiber bundle collapses completely.

Thus with very good approximation we have the simple connection 
\begin{equation}
t_f = 2 t_0.
\end{equation}
When the breaking rate starts increasing we are halfway to complete collapse!\\

\subsection{ Displaced uniform distribution}

Consider a uniform distribution on the interval $(x_l,1)$: 
\begin{equation} 
 p(x) = \left\{ \begin{array}{cc}
\frac{1}{1-x_l}&x_l\leq x\leq 1\\
0 & \mbox{otherwise} 
\end{array}\right\}
\label{displaced}\end{equation}
 
Thus
\begin{equation} 
 P(x) = \left\{ \begin{array}{cc}
0 & x<x_l\\
\frac{x-x_l}{1-x_l} & x_l\leq x \leq 1
\end{array}\right\}
\end{equation}

\vspace{1cm}

Simulations of the breaking rate gives qualitatively the same behavior as for the uniform distribution (Fig.3).\\

\begin{center}
\includegraphics[width=6cm,height=6cm]{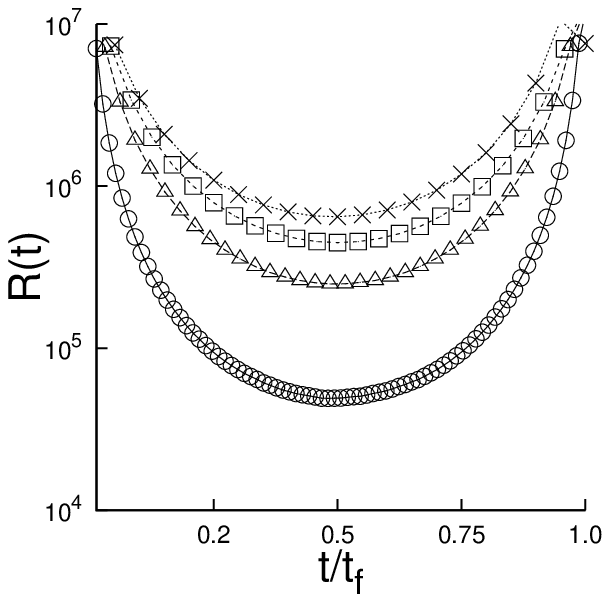}
\par\end{center}

\noindent {\small FIG.\ 3. The breaking rate $R(t)$  vs. the rescaled step 
variable $t_f/t$  for the displaced uniform threshold distribution 
(\ref{displaced}). Here $x_l=0.2$ and $N=5 \times 10^7$. 
Different symbols are used for different excess stress levels 
$\sigma -\sigma _c$: 0.001  (circles), 0.003 (triangles), 0.005 (squares) 
and 0.007 (crosses). }\\{\small \par}

For this distribution eq.(\ref{load}) gives
\begin{equation} 
 \sigma = x(1-P(x)) =\frac{x(1-x)}{1-x_l},
\end{equation}
with a maximum 
\begin{equation} 
\sigma_c =\frac{1}{4(1-x_l)}.\label{sigc}
\end{equation}
at $x=x_c=1/2$.

The iteration (\ref{n}) takes now the form 
\begin{equation} n_{t+1} = \frac{1-\sigma/n_t}{1-x_l}.\label{Uit}\end{equation}
This can be cast in a familiar form. Introduce
\begin{equation} 
 y_t = n_t \;(1-x_l)
\end{equation}
in (\ref{Uit}) to obtain the iteration for $y_t$:

\begin{equation}
 y_{t+1} = 1-\sigma (1-x_l)\cdot\frac{1}{y_t}.
\end{equation}
By (\ref{sigc}) the critical value of $\sigma (1-x_l)$ is $1/4$, so we may write
\begin{equation}
 \sigma (1-x_l)= {\textstyle \frac{1}{4}} + \epsilon,
\end{equation}
where again $\epsilon$ is assumed to be small and positive. Then we are back to the same iteration (\ref{unin}) as for the usual 
uniform distribution: 
 \begin{equation}
y_t = {\textstyle \frac{1}{2}} - \sqrt{\epsilon}\;\tan\left[\tan^{-1}\left(\frac{\frac{1}{2}-y_0}{\sqrt{\epsilon}}\right)+t\;\tan^{-1}(2\sqrt{\epsilon})\right]
\label{Ures}\end{equation}
or, since $y_0=n_0(1-x_l)=1-x_l$:
\begin{equation}
y_t=\frac{1}{2} -\sqrt{\epsilon}\;\tan\left[-\tan^{-1}((1/2-x_l)/\sqrt{\epsilon})+t\tan^{-1}(2\sqrt{\epsilon})\right]
\end{equation}
For simplicity write this as
\begin{equation} y_t = {\textstyle \frac{1}{2}} -\sqrt{\epsilon}\;\tan (at-b),\label{Y}\end{equation}
with 
\begin{equation}
 a = \tan^{-1}(2\sqrt{\epsilon})\hspace{5mm}\mbox{and}\hspace{5mm}b=\tan^{-1}((1/2-x_l)/\sqrt{\epsilon})
\end{equation}
The breaking rate (treating $t$ as continuous) is
\begin{equation}
 R(t)=-\frac{dn_t}{dt} = -\frac{1}{1-x_l}\frac{dy_t}{dt} =\frac{\sqrt{\epsilon}}{1-x_l} \cos^{-2}(at-b)
\end{equation}

The minimum breaking rate occurs when $dR/dt\propto \sin(2at-2b)=0$, i.e.\ at
$ t_0 = b/a.$ For small $\epsilon$ we use the identity
\begin{equation}
\tan^{-1}(1/\eta) = \pi/2 - \tan^{-1}(\eta),
 \label{tang}\end{equation}                       
  and obtain approximately for small $\epsilon$
\begin{equation}
 a\approx 2\sqrt{\epsilon} \mbox{ and } b\approx \pi/2 - \sqrt{\epsilon}/(1/2-x_l)
\end{equation}
Using this, we obtain to leading order
\begin{equation}
 t_0 \simeq  \frac{\pi}{4\sqrt{\epsilon}}. 
\end{equation}

 A good approximation to the collapse point $t_f$ is obtained by selecting the $t$ for which $n_t$ or $y_t$ vanishes. From (\ref{Y}) we see that this occurs for a $t_f$ given by
\begin{equation}
 {\textstyle \frac{1}{2}} -\sqrt{\epsilon}\;\tan (at_f-b) = 0,
\end{equation}
i.e.
\begin{equation}
 t_f = [b+\tan^{-1}(1/2\sqrt{\epsilon})]/a. 
\end{equation}
Again, by using (\ref{tang}) we have for small $\epsilon$
\begin{equation}
 t_f = \frac{\pi/2-\sqrt{\epsilon}/(1/2-x_l)+\pi/2-2\sqrt{\epsilon}}{2\sqrt{\epsilon}}=\frac{\pi}{2\sqrt{\epsilon}}\;\left(1+{\cal O}(\sqrt{\epsilon})\right).
\end{equation}
Comparing the results for $t_0$ and $t_f$ we have once more 
\begin{equation}
 t_f /t_0 = 2.
\end{equation}
to leading order.\\

\subsection{Weibull distribution}

Let us finally consider a completely different threshold distribution, a
 Weibull distribution of index 5, $P(x)=1-e^{-x^5}$. Simulations reveal that 
the breaking rate has a similar behavior as in the two cases considered 
above (Fig.\ 4). \\

\begin{center}
\includegraphics[width=6cm,height=6cm]{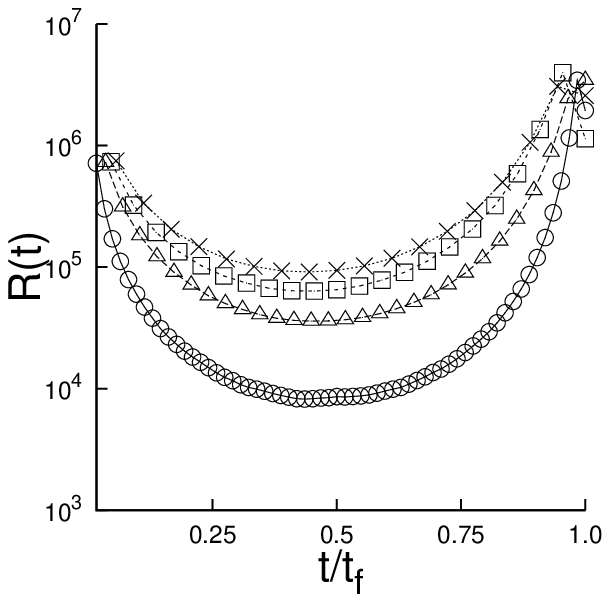}
\par\end{center}
 
\noindent {\small FIG.\ 4. The breaking rate $R(t)$  vs.\ the  rescaled step 
variable $t/t_f$ for a bundle of $N= 10^7$ fibers having a Weibull threshold 
distribution. 
Different symbols are used for different excess stress levels 
$\sigma -\sigma _c$: 0.001  (circles), 0.003 (triangles), 0.005 (squares) 
and 0.007 (crosses). }\\{\small \par}

This case is more complicated, but the analytical ground work has already been 
done in \cite{PH07}. Eq.\ (29) in \cite{PH07} shows that for small $\epsilon$ 
the iteration is of the form
\begin{equation}
n_t = n_c - b\sqrt{\epsilon/C}\; \tan(t\sqrt{C\epsilon}-c).
\label{UW}\end{equation}
Here $n_c=e^{-1/5}, \;C=\frac{5}{2}(5e)^{1/5},\;b=5^{1/5}$, and the constant $c$ 
is determined by the initial condition $n_0=1$:
\begin{equation}
 c= \tan^{-1}\left[(1-n_c)b^{-1}\sqrt{C/\epsilon}\right].
\label{c}\end{equation}
From (\ref{UW}),
the breaking rate equals
\begin{equation}
 R(t) = -\frac{dn_t}{dt} \propto \cos^{-2}(t\sqrt{C\epsilon}-c).
\end{equation}
The breaking rate is a minimum when the cousins takes its maximum value $1$.
This is the case when
\begin{equation}
 t_0 = \frac{c}{\sqrt{C\epsilon}}= (C\epsilon)^{-1/2} \tan^{-1}\left[(1-n_c)b^{-1}\sqrt{C/\epsilon}\right]. 
\end{equation}
The inverse tangent is close to $\pi/2$ when $\epsilon$ is very small. 
Hence, for small overloads, we have in excellent approximation
\begin{equation} 
t_0 = \frac{\pi}{2\sqrt{C\epsilon}}
\label{Wt0}\end{equation}

The collapse point $t_f$ is already evaluated in \cite{PH07}, with the result
\begin{equation}
t_f \simeq \frac{\pi}{\sqrt{C\epsilon}}
\label{Wtf}\end{equation}
for small $\epsilon$ (eq.\ (33) in \cite{PH07}).

Comparison between (\ref{Wtf}) and (\ref{Wt0}) gives 
\begin{equation} 
 t_f \simeq 2 t_0,
\end{equation}
as for the two previous threshold distributions considered. \\

\section {Comments}
We have shown that the complete collapse of fiber bundles occurs at $t_f=2t_0$,
 where $t_0$ denotes the number of steps of the breaking process at which the 
fiber breaking rate has a minimum.  
The results are derived for very small overloads $\epsilon$. For larger overloads the ratio $t_0/t_f$ will not be exactly $0.5$, as illustrated in Fig.\ 5, but nevertheless of the order of 0.5.\\ 
\begin{center}
\includegraphics[width=6cm,height=6cm]{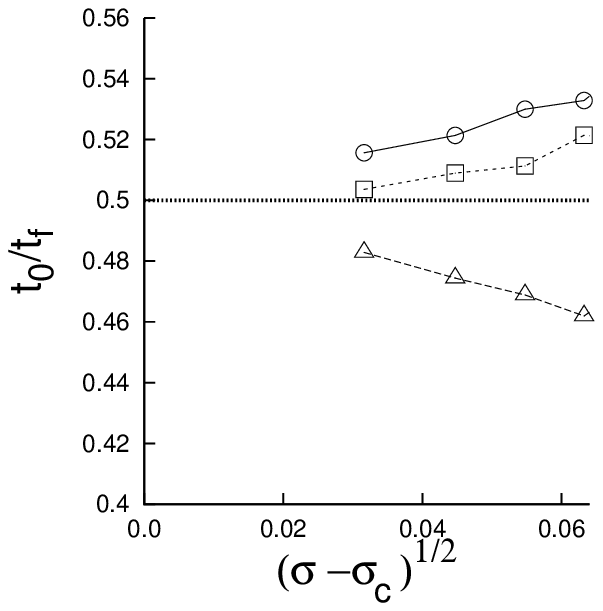}
\par\end{center}
  
\noindent {\small FIG.\ 5. Simulation results for the ratio $t_0/t_f$ vs.\ $(\sigma -\sigma_c) ^{-1/2}$ for  the uniform distribution (circles), the displaced uniform distribution with $x_l=0.2$ (squares) and  for the Weibull distribution (triangles). The graphs are based on $1000$ samples with $N= 10^7$ fibers.}\\{\small \par}

\begin{center}
\includegraphics[width=6cm,height=6cm]{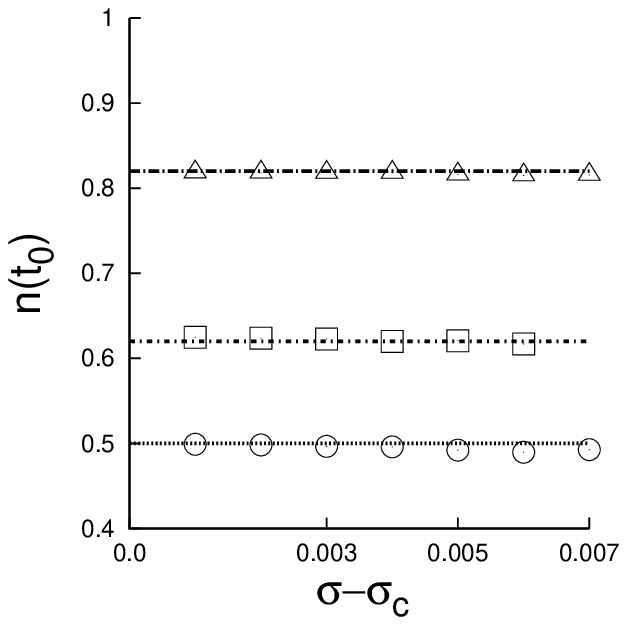}
\par\end{center}
  
\noindent {\small FIG.\ 6. Simulation results for $n(t_0)$ vs.\ $(\sigma -\sigma_c)$ for  the uniform distribution (circles), the displaced uniform distribution with $x_l=0.2$ (squares) and  for the Weibull distribution (triangles). The graphs are based on $1000$ samples with $N= 10^7$ fibers. The straight lines represent the critical value $n_c$ for these three distributions.}\\{\small \par}

Another interesting observation is that at $t=t_0$ the number of unbroken 
fibers in the bundle $n(t_0)$  attains the critical value $n_c$. This can be 
derived analytically by putting the value of $t_0$ in the expressions (10), 
(27) and (37) respectively, for the uniform, the displaced uniform and the 
Weibull 
distribution. The numerical simulations (Fig. 6) strongly support this result.

\section {Summary}

In summary,  we have considered slightly overloaded fiber bundles, and 
investigated how the fiber breaking rate
progresses. It has a minimum after a number of steps $t_0$ of the degradation process, and we have demonstrated that the total bundle collapse occurs near $2t_0$. The demonstration has been performed for three different distributions of fiber thresholds, but the result is doubtlessly universal. Thus the fact that the breaking rate has a minimum predicts not only \textit{that} a global failure will occur, but also estimates \textit{when} it will occur.

\end{document}